# Trust Based Scheme for QoS Assurance in Mobile Ad-Hoc Networks

Sridhar Subramanian[1] and Baskaran Ramachandran[2]

[1]Department of Computer Applications, Easwari Engineering College, Chennai, India.
ssridharmca@yahoo.co.in
[2]Department of Computer Science & Engineering, CEG, Guindy, Anna University, Chennai, India.
baskaran.ramachandran@gmail.com

## ABSTRACT

*A mobile ad-hoc network (MANET) is a peer-to-peer wireless network where nodes can communicate with each other without the use of infrastructure such as access points or base stations. These networks are self-configuring, capable of self-directed operation and hastily deployable. Nodes cooperate to provide connectivity, operates without centralized administration. Nodes are itinerant, topology can be very dynamic and nodes must be able to relay traffic since communicating nodes might be out of range. The dynamic nature of MANET makes network open to attacks and unreliability. Routing is always the most significant part for any networks. Each node should not only work for itself, but should be cooperative with other nodes. Node misbehaviour due to selfish or malicious intention could significantly degrade the performance of MANET. The Qos parameters like PDR, throughput and delay are affected directly due to such misbehaving nodes. We focus on trust management framework, which is intended to cope with misbehaviour problem of node and increase the performance of MANETs. A trust-based system can be used to track this misbehaving of nodes, spot them and isolate them from routing and provide reliability. In this paper a Trust Based Reliable AODV [TBRAODV] protocol is presented which implements a trust value for each node. For every node trust value is calculated and based trust value nodes are allowed to participate in routing or else identified to become a misbehaving node. This enhances reliability in AODV routing and results in increase of PDR, decrease in delay and throughput is maintained. This work is implemented and simulated on NS-2. Based on simulation results, the proposed protocol provides more consistent and reliable data transfer compared with general AODV, if there are misbehaving nodes in the MANET.*

## 1. INTRODUCTION

MANET is a highly challenged network environment due to its special characteristics such as decentralization, dynamic topology and neighbour based routing. They don't rely on existing infrastructure to support communication. Each mobile node acts as an end node when it is the source or destination of a communication and forwards packets for other nodes when it is an intermediate node of the route. Mobile Ad-Hoc network [1] is a system of wireless mobile nodes that self-organizes itself in dynamic and temporary network topologies. Mobile ad hoc networks are suitable for dynamic environment where no infrastructure or temporarily established mobile applications are used, which are cost effective. Ad hoc networks are easier to deploy than wired networks and are found many applications, such as in rescue, battlefields, meeting rooms etc., where either a wired network is unavailable or deploying a wired network is inconvenient. Distributed state in unreliable environment, dynamic topology, limited network capacity, variable link quality, interference and collisions, energy-constrained nodes, flat addressing, scaling issues, heterogeneity are few challenges faced by MANET. Mobile ad hoc network routing protocols face some challenges like node mobility that causes frequent topology changes , the changeable and erratic ability of wireless links and packet losses. Mobile nodes also face troubles like limited power, computing and bandwidth resources.

There have been many ad-hoc routing protocols, which fall into several categories: proactive routing protocols such as dynamic Destination-Sequenced Distance-Vector routing (DSDV), Optimized Link State Routing (OLSR), Topology Broadcast based on Reverse Path Forwarding (TBRPF), on-demand routing protocols such as Dynamic Source Routing (DSR), AODV, Signal Stability-based Adaptive routing (SSA). Proactive routing protocols have little delay for route discovery and are robust enough to link breaks and obtain a global optimal route for each destination. However, their routing overhead is also high. On-demand routing protocols are easy to realize and their overhead is low. But routes in on-demand routing protocols are easy to break in the case of topology variations. In AODV [2] node doesn't have any information about other nodes until a communication is needed. By broadcasting HELLO packets in a regular interval, local connectivity information is maintained by each node. Local connectivity maintains information about all the neighbours.

Recent Qos solutions are planned to operate on trusted environments and totally assume the participating nodes to be cooperative and well behaved [3, 4]. Such assumptions are not valid in dynamic environments like MANETs. Providing different quality of service levels in a persistently changing environment is a challenge because: Unrestricted mobility causes QoS sessions to suffer due to recurrent path breaks, thereby requiring such sessions to be re-established over new paths. The link-specific and state-specific information in the nodes is inherently imprecise due to the dynamically changing topology and channel characteristics. Hence, incorrect routing decisions may chop down Qos parameters performance. Inadequate bandwidth, storage space and battery life also drastically influence the performance of the QoS parameters.

Most security schemes suggested for MANETs tend to build upon some fundamental assumptions regarding the trustworthiness of the participating hosts and the underlying networking system. If MANET is to achieve the same level of acceptance as traditional wired and wireless network infrastructures, then a framework for trust management must become an intrinsic part of its infrastructure. The inherent freedom in self-organized mobile ad hoc networks introduces challenges for trust management, particularly when nodes do not have any prior knowledge of each other. To assure that access to resources is given only to trusted nodes; the trustworthiness among anonymous nodes needs to be formalized. The concept of trust originally derives from social sciences and is defined as the degree of subjective belief about the behaviours of a particular entity [5]. There are four major properties [6, 7] of Trust and they are, Context Dependence where trust relationships are only meaningful in the specific contexts. Function of Uncertainty where trust is an evaluation of probability of if an entity will perform the action. Quantitative Value are where trust can be represented by numeric either continuous or discrete value. Asymmetric Relationship are where trust is the opinion of one entity for another entity.

This traditional AODV is to perform its job based on the trust values calculated for each node and to decide whether to take part or to be isolated from routing. The trust value is calculated for each node based on its success rate and failure rate of transmission. This trust value calculated helps to identify whether the node will be reliable node for performing the routing or may not be reliable for this current transmission. This trust based routing mechanism helps to identify and eliminate misbehaving nodes in MANET and performs an efficient and effective routing. This proposed work also improves the Qos parameters like packet delivery ratio and delay.

## 2. LITERATURE SURVEY

Mobile Ad-Hoc Networks (MANETs) are adaptive and self-organizing, and as a consequence, securing such networks is non-trivial. Mobile ad hoc networks are apt for mobile applications either in antagonistic environments where no infrastructure is available, or temporarily established mobile applications, which are cost decisive. In recent years, application domains of mobile ad hoc networks gain more and more significance in non-military public organizations and in commercial and industrial areas. Medium access control, routing, resource management, quality of service and security are the research areas for mobile ad hoc network. The importance of routing protocols in dynamic networks has directed a lot of mobile efficient ad hoc routing protocols.

A security-enhanced AODV routing protocol called R-AODV (Reliant Ad hoc On-demand Distance Vector Routing) [8] uses a modified trust mechanism known as direct and recommendations trust model and then incorporating it inside AODV. This enhances security by ensuring that data does not go through malicious nodes that have been known to misbehave. Each node is given a trust value and this value is associated with the possibility of the node to perform a packet drop. With the inclusion of trust mechanism, it is expected that using R-AODV would result in a higher percentage of successful data delivery as compared to AODV. It is also expected that the normalized routing load and end-to-end delay would increase.

A framework for estimating the trust between nodes in an ad hoc network based on quality of service parameters using probabilities of transit time variation, deleted, multiplied and inserted packets, processing delays to estimate and update trust [9]. This paper clearly shows that only two end nodes need to be concerned and attain reduced overhead. The framework proposed in this paper is applicable and useful to estimate trust in covert unobservable and anonymous communications. This results in detecting regular packets drops and delay detection.

A schema is formed via direct and indirect approach to compute trust value among anonymous nodes [10]. To evaluate trust values the parameters like reputation, knowledge, observation and context were used. The trust schema that is build is used to allow resource to be shared among trusted nodes. The result obtained is then mapped with the access privileges to take appropriate actions.

A routing protocol, which adds a field in request packet and also stores trust value indicating node trust on neighbour based on level of trust factor [11 is discussed here. The routing information will be transmitted depending upon highest trust value among all. This not only saves the node's power by avoiding unnecessary transmitting control information but also in terms of bandwidth (channel utilization), which is very important in case of MANET. The malicious node can attack on the control packet and misbehave in the network. A trusted path is used irrespective of shortest or longest path, which can be used for communication in the network. It calculates route trust value on the complete reply path, which can be utilized by source node for next forthcoming communication in the network. Thus security level is improved and also malicious node attacks are prevented in the network.

A trust model introduced in the network layer leads to a secure route between source and destination without any intruders or malicious nodes in the network [12]. This trust based routing protocol concentrates both in route and node trust. Node Trust Calculation Process is done by introducing a new data structure neighbour table in each node of the MANET. Node trust is calculated by the collective opinion of node's neighbours. The resultant trust value is placed in trust value field of neighbour table. Node trust calculated based upon the information that one node could collect about the other nodes. Route Trust Calculation Process is done using a modified extended route table. With this minimum overhead, eliminates the malicious node as well as establish a best-trusted route between source and destination.

TAODV [13], an enhanced AODV protocol was proposed with a concept of trust values for calculating trust values of nodes. The changes made to the existing protocol are, two new control packets TREQ (Trust request) & TREP (Trust Reply) and a modified extended routing table with four new fields; positive events, negative events, route status, opinion. This provided a reliable routing

## 3. PROPOSED WORK

Many trust management schemes have been proposed to evaluate trust values and most of the trust-based protocols for secure routing calculated trust values based on the characteristics of nodes behaving properly at the network layer. Trust measurement can be application dependent and will be different based on the design goals of proposed schemes [14]. The trust management metrics include overhead (e.g., control packet overheads), throughput, packet delivery ratio, packet dropping rate, and delay.

Routing in mobile ad hoc networks is pretentious due to the dynamic nature of nodes, which are not stable and keep moving. But still nodes communicate with each other and exchange data within the available nodes on the network. The architecture of the proposed work is presented in figure 1. The node trust plays a very crucial role in MANET routing. Trust factor here focuses on identifying the nodes which not suitable for reliable routing and helps to select an alternate path to carry on routing successfully using reliable nodes. The proposed work concentrates on identifying these unreliable nodes using the trust level values calculated for each node.

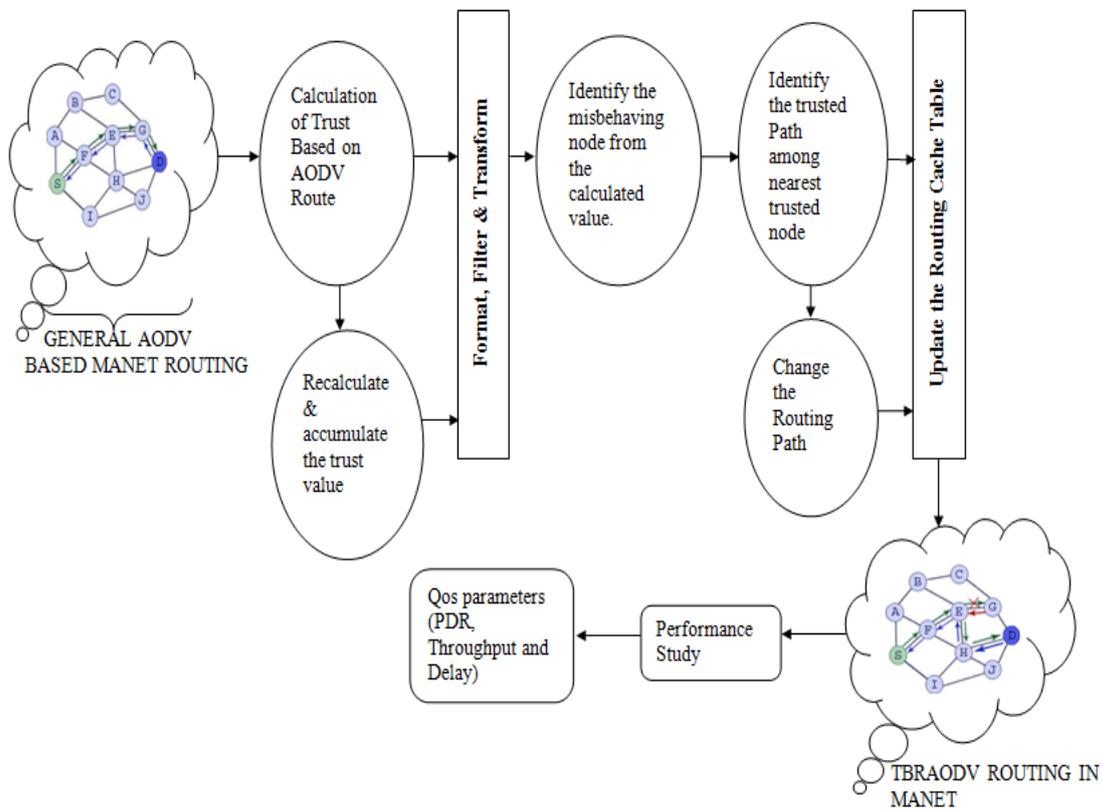

Figure 1. Architecture of proposed TBRAODV routing in MANET

The trust level value calculation is based on the parameters shown in the table 1. The count field describes about two criteria success and failure which describes whether the transmission was a successful transmission or a failure. RREQ and RREP are the route request and route reply respectively which are exchanged between nodes in the network. Data refers to the payload transmitted by the node in the routing path.

Table 1. Trust value calculation parameters

| Count Type | RREQ | RREP | Data |
|---|---|---|---|
| Success | Qrs | Qps | Qds |
| Failure | Qrf | Qpf | Qdf |

The parameter $q_{rs}$ is defined as the query request success rate which is calculated based on number of neighbouring nodes who have successfully received (rreq) from the source node which has broadcasted it, $q_{rf}$ defined as the query request failure rate which is calculated based on number of neighbouring nodes which have not received the query request, $q_{ps}$ is defines as the query reply success rate which is calculated as successful replies (rrep) received by the source node which has sent the rreq and $q_{pf}$ is defined as the query reply failure rate which is calculated based on the number of neighbouring nodes which have not sent the replies for the query request received. $q_{ds}$ is defined as the data success rate calculated based on successfully transmitted data and $q_{df}$ is defined as data failure rate calculated based on data which have failed to reach destination. However, it is known that for every network there will be minimum data loss due to various constraints.

$$Qr = \frac{q_{rs} - q_{rf}}{q_{rs} + q_{rf}} \quad (1)$$

$$Qp = \frac{q_{ps} - q_{pf}}{q_{ps} + q_{pf}} \quad (2)$$

$$Qd = \frac{q_{ds} - q_{df}}{q_{ds} + q_{df}} \quad (3)$$

Where Qr, Qp and Qd are intermediate values that are used to calculate the nodes Request rate, Reply rate and Data transmission rate. The values of Qr, Qp, and Qd are normalized to fall in range of -1 to +1. If the values fall beyond the normalized range then it clearly shows that the failure rate of the node is high and denotes that the corresponding node may not be suitable for routing.

$$TL = T(RREQ)*Qr + T(RREP)*Qp + T(DATA)*Q_d \quad (4)$$

Where, TL is the trust level value and T(RREQ), T(RREP) and T(DATA) are time factorial at which route request, route reply and data are sent by the node respectively. Apart from the above mentioned normalised range, using the above formula the trust level value (TL) is calculated for each node during routing and is checked against the threshold value (assumed to be as 5). If lesser than threshold then there is a possibility for this node to be marked as misbehaving node for the current transmission and will not be suitable for further routing and an alternate path is selected for routing. However, this node may be the best node for some other transmission between some other source and destination in the same network. TBRAODV checks every node with its trust value to make itself robust and trustworthy for effective and efficient routing and also to assure qos in MANET.

For the sample network shown in figure 2, the path selected is S →F →E →G →D. For example, Node E has four neighbours and for this node the trust value calculation is to be done.

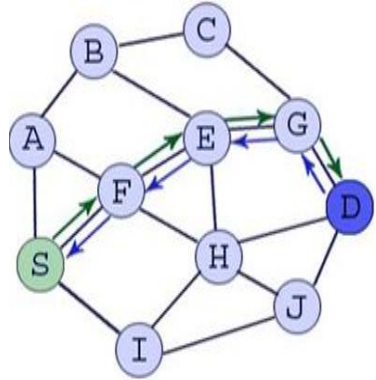

Figure 2. Sample network to implement TBRAODV

For node E the trust value calculation table is given in table 2 which contains the success and failure rate of route request, reply and data.

Table 2. Trust value calculation for Node E

| Count Type | RREQ | RREP | Data |
|---|---|---|---|
| Success | 4 | 4 | 900 |
| Failure | 0 | 0 | 100 |

$Qr = ( 4 - 0 ) / ( 4 + 0 ) = 1$

$Qp = ( 4 - 0 ) / ( 4 + 0 ) = 1$

$Qd = ( 900-100 ) / ( 900+100) = 0.8$

The values of Qr, Qp, and Qd are falling within the normalized range fixed (i.e) -1 to +1. Thus the trust value is calculated for the node E.

$TL = 1*1!+1*2!+0.8*3! = 7.8$ (which is more than 5) thus making this node a reliable node for routing. This trust calculation is done for all nodes in the routing path to monitor nodes behaviour. If the failure rate increases it automatically affects the Qr, Qp and Qd values thus making them fall beyond the normalized values thus resulting in trust value less than the threshold.

## 4. EVALUATION RESULTS

The performance of proposed TBRAODV protocol is analyzed using NS-2 simulator. The network is designed using network simulator with maximum of 50 nodes. Other parameters based on which the network is created are given in Table 3. Results are obtained from this simulation applying both general AODV and proposed TBRAODV protocols. The proposed TBRAODV protocol has shown good improvement over the QoS parameters like PDR & Delay. PDR is increased and delay is reduced compared to the general AODV. Throughput is maintained. Graphs are used to compare the results of the existing AODV and proposed TBRAODV protocol and clearly indicate the improvement of the proposed protocol.

Table 3. Simulation Parameter Values

| Parameter | Value |
|---|---|
| Network size | 1600 x 1600 |
| Number of nodes | 50 |
| Transmission range | 250 meters. |
| Movement speed | 100 kbps |
| Traffic type | CBR |
| Packet size | 5000 |
| Simulation time | 30 minutes. |
| Maximum speed | 100 kbps |
| Time interval | 0.01 sec. |
| MAC layer protocol | IEEE 802.11 |
| Protocol | AODV |
| NS2 version | 2.34 |

Simulation results were obtained and compared. The results show a good improvement than the exiting approach. The proposed protocol has performed well than the existing AODV protocol which lacks in Qos parameters like PDR and delay when compared with the proposed TBRAODV protocol. The results obtained are shown in Table 4, which shows the values obtained using general AODV and proposed TBRAODV at different node sizes. The traditional AODV is affected due to the existence of misbehaving nodes, which results in low packet delivery ratio and also causes the delay to increase. The proposed protocol has shown improved Qos parameters values where trust values are used to identify the misbehaving nodes in the route and immediately take an alternate path to successfully complete the routing. This approach of the proposed TBRAODV protocol has resulted in an increased packet delivery ratio and a decreased delay involved in routing.

Table 4. Comparison of Result with node size

| Node Size | General AODV | | | Proposed TBRAODV | | |
|---|---|---|---|---|---|---|
| | PDR | Delay | Throughput | PDR | Delay | Throughput |
| 25 | 82.98 | 0.24615 | 75771.43 | 92.20 | 0.22153 | 75771.43 |
| 50 | 70.05 | 0.84972 | 114559.89 | 91.06 | 0.64979 | 114559.89 |
| 100 | 64.43 | 1.44347 | 148339.67 | 90.03 | 0.92683 | 148339.67 |
| 200 | 62.36 | 1.65589 | 150748.56 | 84.32 | 0.93536 | 150748.56 |
| 300 | 60.65 | 1.78687 | 150836.74 | 81.26 | 0.94825 | 150836.74 |

Figure 3 indicates how the proposed TBRAODV protocol has shown a good decrease in Delay when compared to the general AODV. Figure 4 shows the increase in PDR when compared with the general AODV.

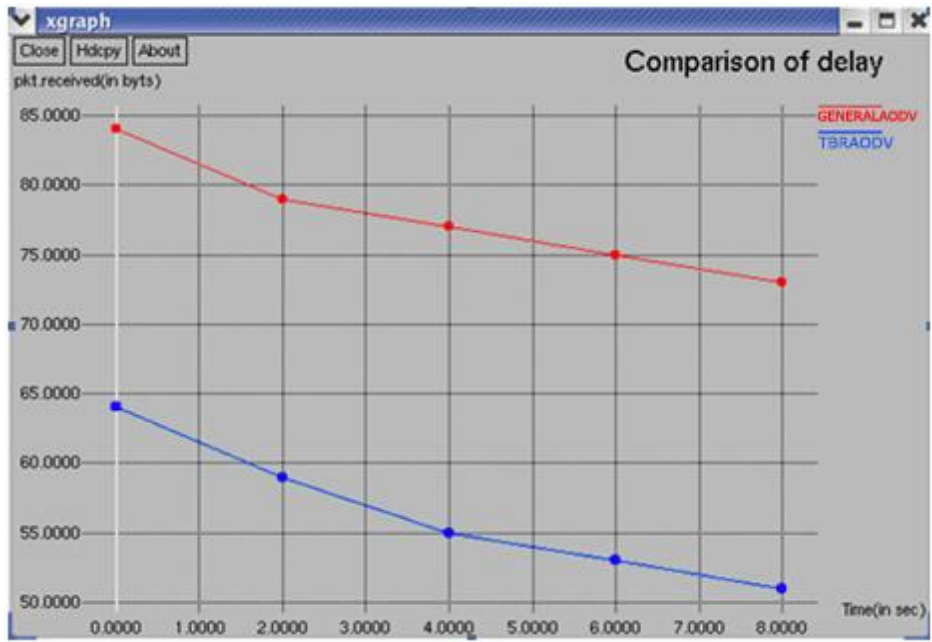

Figure 3. Comparison of general AODV Delay and TBRAODV Delay

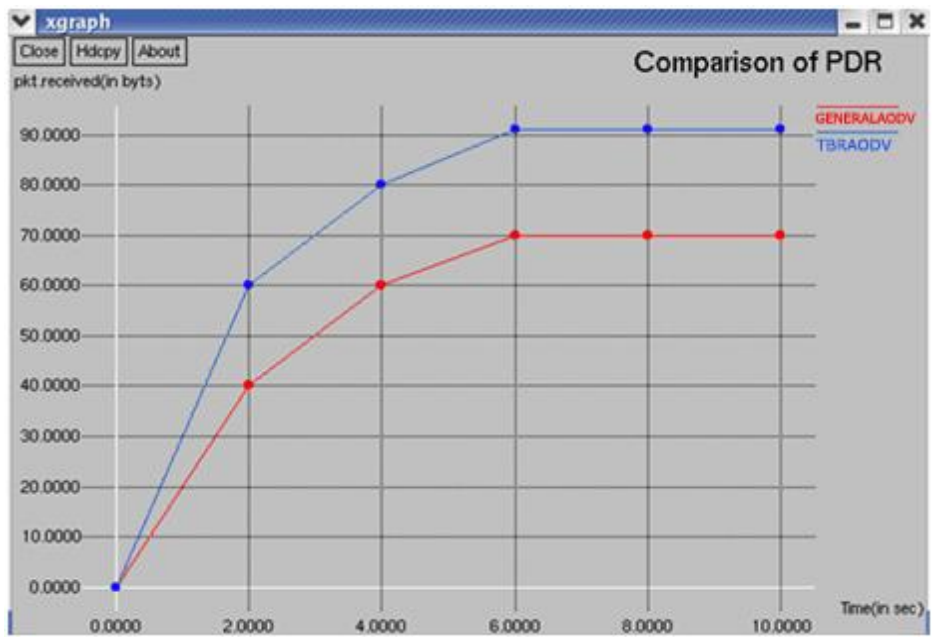

Figure 4 . Comparison of general AODV PDR and TBRAODV PDR

## 5  CONCLUSION AND FUTURE ENHANCEMENTS

In this paper, a trust based reliable protocol TBRAODV is proposed. Trust level values for each node are calculated to identify the misbehaving nodes during routing. If node is misbehaving then it leads to an alternate path selection for further reliable routing. This trust based routing mechanism has proved to be increasing the performance of the proposed TBRAODV protocol and also shows good improvement of Qos parameters like PDR and delay. Rather implementing reliability with trust alone some energy constraints on each node along with trust schemes for a node will provide better reliability for MANET routing. The same scheme can also be implemented on other MANET routing protocols and also implement some techniques for authenticating the packet and the node which take part in routing.

## REFERENCES

[1] Kortuem.G., Schneider. J., Preuitt.D, Thompson .T.G.C, F'ickas.S. Segall.Z. "When Peer to-Peer comes Face-to-Face: Collaborative Peer-to-Peer Computing in Mobile Ad hoc Networks", 1st International Conference on Peer-to-Peer Computing, August, Linkoping, Sweden, pp. 75-91 (2001)

[2] C.Perkins, E.Royer and S.Das, "Ad hoc on-demand Distance Vector Routing", RFC-3651

[3] Hu, Y., "Enabling Secure High-Performance Wireless Ad Hoc Networking," PhD Thesis, Carnegie Mellon University (CMU), (2003)

[4] IIyas M., The Handbook Of Wireless Ad Hoc Network, CRC, (2003)

[5] K. S. Cook (editor), Trust in Society, vol. 2, Feb. 2003, Russell Sage Foundation Series on Trust, New York

[6] Farooq Anjum, Dhanant Subhadrabandhu and Saswati Sarkar "Signature based Intrusion Detection for Wireless Ad-Hoc Networks: A Comparative study of various routing protocols" in proceedings of IEEE 58th Conference on Vehicular Technology, 2003.

[7] Marc Branchaud, Scott Flinn,"x Trust: A Scalable Trust Management Infrastructure"

[8] Hothefa Sh.Jassim, Salman Yussof, "A Routing Protocol based on Trusted and shortest Path selection for Mobile Ad hoc Network", IEEE 9th Malaysia International Conference on Communications (2009)

[9] D. Umuhoza, J.I. Agbinya., "Estimation of Trust Metrics for MANET Using QoS Parameter and Source Routing Algorithms", The 2nd International Conference on Wireless Broadband and Ultra Wideband Communications (2007)

[10] Asmidar Abu Bakar, Roslan Ismail, Jamilin Jais, "Forming Trust in Mobile Ad -Hoc Network", 2009 International Conference on Communications and Mobile Computing (2009)

[11] R. S. Mangrulkar, Dr. Mohammad Atique, "Trust Based Secured Adhoc on Demand Distance Vector Routing Protocol for Mobile Adhoc Network" (2010)

[12] A.Menaka Pushpa M.E., "Trust Based Secure Routing in AODV Routing Protocol" (2009)

[13] "TAODV: A Trusted AODV Routing protocol for Mobile ad hoc networks" (2009)

[14] A Survey on Trust Management for Mobile Ad Hoc Networks Jin-Hee Cho, Member, IEEE, Ananthram Swami, Fellow, IEEE, and Ing-Ray Chen, Member, IEEE

[15] Sridhar, S., Baskaran, R.: Conviction Scheme for Classifying Misbehaving Nodes in Mobile Ad Hoc Networks in the proceedings of CCSIT 2012 published by Springer (LNICST) 2012